\documentclass[twocolumn,runningheads,natbib]{svjour2}
\smartqed  
\usepackage{graphicx}
\usepackage{mathptmx}      
\journalname{Astrophysics and Space Science}
\begin{document}

\title{RX J1856.5-3754 as a possible Strange Star candidate.
\thanks{This work was patially supported by PAPIIT, UNAM, grant IN119306.
        J.A.H. studies at UNAM and travel to London are covered by
        fellowships from UNAM's Direcci\'on General de Estudios de Posgrado.}
}

\author{Jillian Anne Henderson         \and
        Dany Page}

\institute{J. A. Henderson \and D. Page \at
              Instituto de Astronom\'ia \\
	      Universidad Nacional Aut\'onoma de M\'exico\\
	      Ciudad Universitaria\\
	      M\'exico, D.F., CP 04510 \\
              \email{hendersj@astroscu.unam.mx}\\ 
	      \email{page@astroscu.unam.mx} 
}

\date{Received: date / Accepted: date}

\maketitle

\begin{abstract}

RX J1856.5-3754 has been proposed as a  stra\-nge star candidate due to 
its very small apparent radius measured from its X-ray thermal spectrum. 
However, its optical emission requires a much larger radius and thus most 
of the stellar surface must be cold and undetectable in X-rays. 
In the case the star is a neutron star such a surface temperature 
distribution can be explained by the presence of a strong toroidal field 
in the crust \citep{PMP06,GKP06}.
We consider a similar scenario for a strange star with a thin baryonic 
crust to determine if such a magnetic field induced effect is still 
possible. 

\end{abstract}

\section{Introduction}
\label{sec:1}

Quark stars have long ago been proposed as an alternative 
to neutron stars \citep{I70}.
The ``strange matter hypothesis'' \citep{W84} gave a more precise
theoretical formulation for their existence: 
that at zero pressure
three flavor quark matter, i.e. with $u$, $d$ and $s$ quarks, has
a lower density per baryon than nuclear matter and would hence
be the true ground state of baryonic matter.
These stars are now called ``strange stars'' \citep{AFO86,HZS86}
and share many similarities with neutron stars:
they can have similar masses, have similar radii in the observed
range of masses, similar cooling histories and, to date, it has been
practically impossible to conclusively prove or disprove their existence
(for recent reviews, see \citealt{W05,PGW06,PR06}).

One possible distinctive property of a strange star could be a small radius.
Given the impossibility to treat quark-quark interactions from first
principles, i.e. starting from 
Q.\-C.\-D., at densities relevant for compact
stars, only simplified models are possible and results are, naturally,
model dependent.
However, several classes of such models do predict  small radii, $5 < R < 10$ km 
at masses $\sim 1.4 M_\odot$ (see, e.g., \citealt{Detal98,Hetal01}),
and {\em all} models predict very small radii, $\leq 5$ km, at masses 
$\ll 1 M_\odot$.
Hence, measurement of a compact star radius giving a radius $\ll 10$ km
directly allows a claim for a strange star candidate.

The ``Magnificent Seven'' \citep{H06} arouse great expectations to measure
compact star radii with high enough accuracy to put strong constraints on
the dense matter equation of state.
In particular, fits of the observed soft X-ray thermal spectrum of
RX J1856.5-3754 \citep{Petal02} pointed to a very small radius
and lead to the claim that this object may be a strange star \citep{Detal02}.
However, observations in the optical band allowed the identification of the
Rayleigh-Jeans tail of a second component of the surface thermal emission,
corresponding to a lower temperature and much larger radius than the
component detected in the X-ray band.
An interpretation of these results is that the surface temperature of the
star is highly non-uniform \citep{Petal02,Tetal04}, possibly due to the
presence of a strong magnetic field.
Models of surface temperature distribution with purely poloidal magnetic
fields \citep{P95,GKP04} do predict non-uniform surface temperature
distributions, , but such inhomogeneities are not strong enough to produce such small
X-ray emitting regions surrounded by large cold regions detectable in
the optical band as observed.
However, inclusion of a toroidal component of the magnetic field, confined to
the neutron star crust, has a dramatic effect 
(\citealt{PMP06,GKP06}; see also \citealt{Pons06,P06}):
this field component inhibits heat from the stellar core to flow
to the surface through most of the crust, except for small domains
surrounding the magnetic axis, and results in highly non-uniform surface
temperature distributions producing good fits to the observed
thermal spectra, from the optical up to the X-ray band.

These models of small hot regions, detected in the X-ray band,
surrounded by large cold regions, detected in the optical band, which allow
to reproduce the entire observed thermal spectrum and results in large radii 
for the star are in contradiction with the proposed strange star
interpretation of RX J1856.5-3754, which was based on the small radius 
detected in the X-ray band.
Here we want to push the discussion one step further: are these highly
non-uniform surface temperature distributions, assuming they are real,
incompatible with a strange star model ?
We consider strange stars having a thin crust, composed of normal bayonic
matter, with a strong magnetic field. 
Since a strange star crust can, at most, reach the neutron drip density,
it is much thiner than the crust of a normal neutron star and
the specific question is: 
can such a thin layer produce the surface temperature distributions
deduced from observation ?

\section{The Strange Star Models}
\label{sec:2}

\begin{figure}
\centering
  \includegraphics[width=0.45\textwidth]{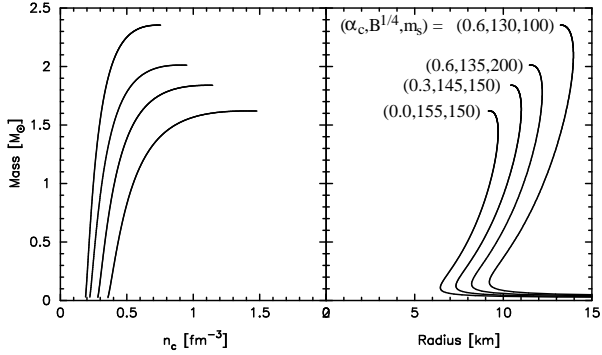}
\caption{Mass vs. central density (left) and radius (right)
for strange stars built with four MIT-bag inspired models of 
strange matter (model parameters are indicated in the right panel)
and covered by a thin baryonic crust (see Figure \protect\ref{Fig:2}).
}
\label{Fig:1}
\end{figure}

We will consider strange star models built on the MIT-bag inspired
equation of state of \cite{FJ84} which has three parameters:
the QCD coupling constant $\alpha_s$, the bag constant $B$ and the
strange quark mass $m_s$ ($u$ and $d$ quarks are treated as massless).
Figure~\ref{Fig:1} illustrates four families of such strange matter models: 
by varying the parameters, these equations of state allow the production of
a wide range of models, from very compact stars up to very large ones.
It is important to notice from this figure that, depending on the assumed
parameters of the model, strange stars can have large radii and thus,
although a small radius measurement is a strong argument in favor of a
strange star, a large radius is {\em not} an argument against a stange star.

\begin{figure}
\centering
  \includegraphics[width=0.35\textwidth]{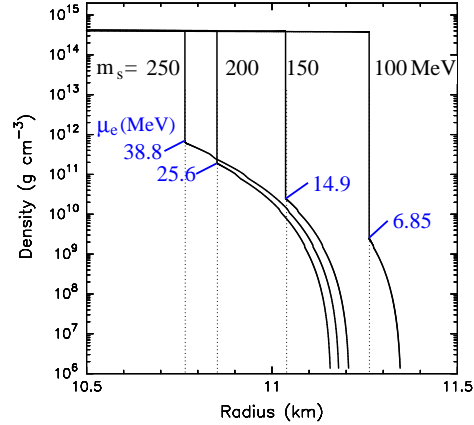}
\caption{Density profile of the upper layers
of four strange stars built with four MIT-bag inspired models of 
strange matter, using $\alpha_s = 0.3$, $B^{1/4} = 140$ MeV and
four different strange quark masses as indicated.
Indicated are also the electron chemical potentials $\mu_e$ at the
quark surface, which is the parameter determining the maximum baryonic
crust density, i.e., the baryonic crust thickness.
The crust equation of state is from \cite{HZD89}.
}
\label{Fig:2}
\end{figure}

On top of the quark matter, a thin crust can exist as long as the
electron density within it is smaller than that in the quark matter
\citep{AFO86}.
Such a baryonic crust is, however, much thinner than a neutron star crust as
illustrated in Figure~\ref{Fig:2}.

\begin{figure}
\centering
  \includegraphics[width=0.40\textwidth]{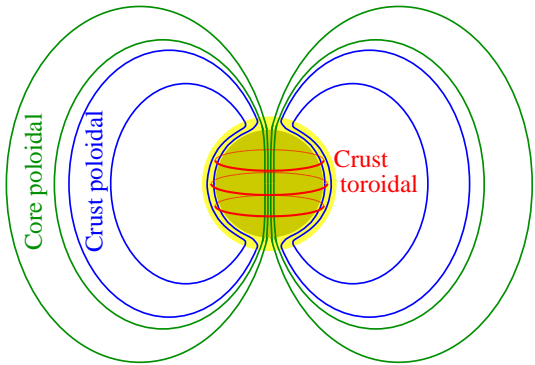}
\caption{The three components of the magnetic field considered in this work}
\label{Fig:3}
\end{figure}

Following the neutron star models presented by \cite{GKP06} we consider
dipolar magnetic fields with three components (Figure~\ref{Fig:3}):
a poloidal one maintained by currents in the quark core,
$\mathbf{B}^\mathrm{core}$,
a poloidal one maintained by currents in the baryonic crust,
$\mathbf{B}^\mathrm{crust}$,
and a toroidal one maintained by currents in the baryonic crust,
$\mathbf{B}^\mathrm{tor}$.
The separation between currents in the crust and in the core is
motivated by the likely fact that quark matter forms a Maxwell
superconductor \citep{A01,PR06}:
at the moment of the phase transition, occurring very early in the
life of the star, superconductivity will prevent any current in the crust 
from penetrating the core while currents in the core will become
supercurrents and be frozen there.
Moreover, flux expulsion due to the star's spin-down can also significantly
increase the crustal field at the expense of the core field.

The importance of the crustal field is that its field lines are forced
to be closed within the crust and hence it has a very large
meridional component $B_\theta$, compared to the core component.
Due to the classical Larmor rotation of electrons, a magnetic field 
causes anisotropy of the heat flux and the heat conductivity becomes 
a tensor whose components perpendicular, $\kappa_\perp$, and parallel,
$\kappa_\|$, to the field lines become
\begin{equation}
\kappa_\perp = \frac{\kappa_0}{1+(\Omega_B \tau)^2}
\;\;\;\;\;
\mathrm{and}
\;\;\;\;\;
\kappa_\| = \kappa_0
\label{Equ:kappa}
\end{equation}
where $\kappa_0$ is the conductivity in the absence of a magnetic field,
$\Omega_B$ the electrons cyclotron frequency and $\tau$ their collisional 
relaxation time.
The large values of $B_\theta$ in the crust have the effect of inhibiting
radial heat flow except in regions close to the magnetic axis where
$B_r$ dominates over $B_\theta$ \citep{GKP04}.

\section{Strange Stars - Results}
\label{sec:3}

We performed heat transport calculations using the 2D code described in
\cite{GKP04} which incorporates the thermal conductivity anisotropy
descri\-bed by Eq.~\ref{Equ:kappa}.
Details of the crust microphysics are as described in \cite{GKP04,GKP06}.
We consider a strange star model with a radius of $\sim$ 11 km and a 
baryonic crust of thickness $\sim$ 250 m for a 1.4 $M_\odot$ mass.
The two poloidal components of the magnetic field are parametrized
by $B_o^\mathrm{core}$ and $B_o^\mathrm{crust}$ which are the strengths
of the corresponding field components at the surface of the star along the
magnetic axis so that, ideally, $B_o^\mathrm{core} +B_o^\mathrm{crust}$ 
would be the dipolar field estimated from the star's spin-down.
Notice that the maximum value of $B^\mathrm{core}$ in the crust is only
slightly larger than $B_o^\mathrm{core}$ while {\bf maximum values of }
$\mathbf{B^\mathrm{crust}}$ {\bf are up to almost 100 times larger than} 
$\mathbf{B_o^\mathrm{crust}}$ due to its
large tangential component, $B_\theta^\mathrm{crust}$, resulting from the 
compression of the field within the narrow crust.
The strength of the toroidal field is parame\-tri\-zed by $B_o^\mathrm{tor}$, 
defined as the maximum value reached by $B^\mathrm{tor}$ within the crust.
We keep $B_o^\mathrm{core}$ at $10^{13}$ G and vary 
$B_o^\mathrm{crust}$ and $B_o^\mathrm{tor}$.

\begin{figure*}
\centering
  \includegraphics[width=0.75\textwidth]{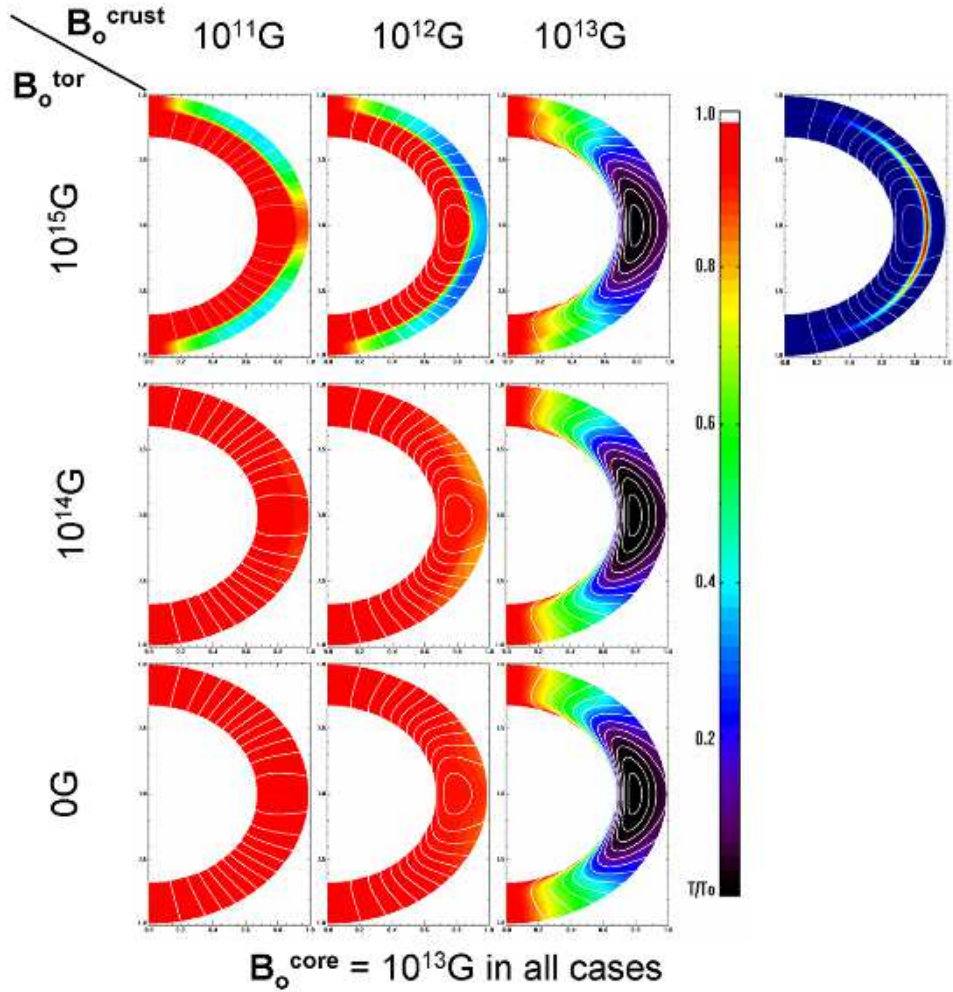}
\caption{Crustal temperature profiles for various strange star magnetic 
field structures according to the model parameters (see text for details). 
The radial aspect of the crust has been stretched by a factor of 15 
in order to clearly show the thermal structure,
The upper right profile shows the field lines of the poloidal component
$\mathbf{B}^\mathrm{core}+\mathbf{B}^\mathrm{crust}$ (white lines) 
and the intensity distribution of the toroidal component $\mathbf{B}^\mathrm{tor}$ (in colours).}
\label{Fig:4}
\end{figure*}

We display in Figure~\ref{Fig:4} the resulting crustal temperature 
profiles for several typical values of $B_o^\mathrm{crust}$ and 
$B_o^\mathrm{tor}$. 
One sees that, independently of the strength of the poloidal component,
$B_o^\mathrm{tor}$ needs to reach $10^{15}$ G to have a significant
effect, a result similar to what was obtained by \cite{GKP06}
for the neutron star case.
However, independently of the value of $B_o^\mathrm{tor}$, once
$B_o^\mathrm{crust}$ reaches $10^{13}$ G highly non-uniform temperature
profiles develop in the thin strange star crust:
such profiles are sufficiently non-uniform to produce the wanted surface 
temperature distribution, i.e. small hot regions surrounded by extended
cold ones.

\section{Discussion and Conclusions}
\label{sec:4}

The optical + X-ray spectrum of RX J1856.5-3754, when fitted with
blackbodies, requires two components with very different temperatures
and emitting areas. 
The implied highly non-uniform surface temperature distribution can
be physically justified by the introduction of a very strong magnetic
field, whose supporting currents are mostly located whithin the star's
crust.
We have shown here that, similarly to the neutron star case, such field
configurations can be found in the case of a strange star, despite the
shallowness of its baryonic crust.
However, the strength of these fields, either the toroidal component
$\mathbf{B}^\mathrm{tor}$ or the crust anchored poloidal one $\mathbf{B}^\mathrm{crust}$
must reach strengths close to, or above, $10^{15}$ G to produce the
desired temperature anisotropy.
Similarly to the neutron star case, such surface temperature distributions
impose severe, but not unrealistic, restrictions on the orientation of
either the observer or the magnetic field symmetry axis with respect
to the rotation axis to explain the absence of pulsations
\citep{BR02,GKP06}.

That the thin stange star crust can support such huge field strenghs
is an open question, but a positive answer seems doubtful as some simple
estimates illustrate.
The magnetic shear stress, $B_r B_\theta/4\pi$, reaches 
$10^{26}$ dyne cm$^{-2}$ in the crust when $B_o^\mathrm{crust}$ reaches
$10^{13}$ G, which is comparable or higher to the maximum value 
sustainable by a crust of thickness $\Delta \sim 200-300$ m
\citep{R04}:
violent reajustments of the crust are expected but have yet to be observed
in any of the ``Magnificent Seven''.
Moreover, the ohmic decay time in the low density crust is relatively short,
less than $10^5$ yrs \citep{PGZ00}, and the presence of such a strong field
in a $\sim 10^6$ yrs old star would require an initial crustal field about
two orders of magnitude higher when the star was young.
(However, the highly non-linear evolution of coupled strong poloidal and
toroidal magnetic fields remains to be studied under such condictions).

We have also adopted the very ingenuous assumption that the surface emits
as a perfect blackbody.
A condensed matter surface may simulate a blackbody spectrum
\citep{TZD04,PMP05,vA05} but such models still require strong fields
to produce a non-uniform temperature distribution \citep{PMP06}.
However, other interpretations are possible, such as a 
thin atmosphere atop a solid surface \citep{MZH03,Ho06} which may be able to
reproduce both the optical and X-ray spectra without invoking strongly
non-uniform temperatures and can be applied 
as well to strange stars with a crust as to neutron stars since they
only consider the very surface of the star.

In conclusion, the crustal field scenario, which is successfull when
applied to neutron star models in order to explain the observed thermal
spectrum properties of RX J1856\-.5-3754, can be ``successfully'' applied
to a strange star mo\-del but requires such a huge magnetic field confined 
within such a thin crust that its applicability is doubtful.
It is hence difficult to conciliate the observed, from optical to X-ray, properties
of  RX J1856.5-3754 with a strange star interpretation unless these are due
exclusively to the emitting properties of its surface.


\begin{acknowledgements}
We thank M. K\"uker and U. Geppert for allowing us to use
their 2D transport code.
\end{acknowledgements}




\begin{thebibliography}{3}

\bibitem[\protect\citeauthoryear{Alcock et al.}{1986}]{AFO86}
Alcock, C., Farhi, E., Olinto, A.
ApJ \textbf{310}, 261 (1986)

\bibitem[\protect\citeauthoryear{Alford}{2001}]{A01}
Alford, M.
Annu. Rev. Nucl. Part. Sci. \textbf{51}, 131 (2001)

\bibitem[\protect\citeauthoryear{Braje and Romani}{2002}]{BR02}
Braje, T.M., Romani, R.W.
ApJ \textbf{580}, 1043 (2002)

\bibitem[\protect\citeauthoryear{Dey et al.}{1998}]{Detal98}
Dey, M., Bombaci, I., Dey, J. et al.
Phys. Lett. \textbf{B438}, 123 (1998)

\bibitem[\protect\citeauthoryear{Drake et al.}{2002}]{Detal02}
Drake, J.J., Marshall, H.L., Dreizler, S. et al..
ApJ \textbf{572}, 996 (2002) 

\bibitem[\protect\citeauthoryear{Farhi and Jaffe}{1984}]{FJ84}
Farhi, E., Jaffe, R.L.
Phys. Rev. {\bf D30}, 2379 (1984)

\bibitem[\protect\citeauthoryear{Geppert et al.}{2004}]{GKP04}
Geppert, U., K\"uker, M., Page, D.
A\&A {\bf 426}, 267 (2004)

\bibitem[\protect\citeauthoryear{Geppert et al.}{2006}]{GKP06}
Geppert, U., K\"uker, M., Page, D.
A\&A \textbf{457}, 937 (2006)

\bibitem[\protect\citeauthoryear{Haberl}{2007}]{H06}
Haberl, F.
in these proceedings (2007)

\bibitem[\protect\citeauthoryear{Haensel et al.}{1986}]{HZS86}
Haensel, P., Zdunik, J.L., Schaeffer, R.
A\&A \textbf{160}, 121 (1986)

\bibitem[\protect\citeauthoryear{Haensel et al.}{1989}]{HZD89}
Haensel, P., Zdunik, J.L., Dobaczewski, J.
A\&A \textbf{222}, 353 (1989)

\bibitem[\protect\citeauthoryear{Hanauske et al.}{2001}]{Hetal01}
Hanauske, M., Satarov, L.M., Mishustin, I.N. et al.
Phys. Rev. \textbf{D64}. 3005 (2001)

\bibitem[\protect\citeauthoryear{Ho et al.}{2007}]{Ho06}
Ho, W.C.G., Kaplan, D.L., Chang, P. et al.
in these proceedings (2007)

\bibitem[\protect\citeauthoryear{Itoh}{1970}]{I70}
Itoh, N.
Prog. Theor. Phys. \textbf{44}, 291 (1970)

\bibitem[\protect\citeauthoryear{Motch et al.}{2003}]{MZH03}
Motch, C., Zavlin, V.E., Haberl, F.
A\&A \textbf{408}, 323 (2003)

\bibitem[\protect\citeauthoryear{Page}{1995}]{P95}
Page, D.
ApJ \textbf{442}, 273 (1995)

\bibitem[\protect\citeauthoryear{Page}{2007}]{P06}
Page, D.
in these proceedings (2007)

\bibitem[\protect\citeauthoryear{Page et al.}{2000}]{PGZ00}
Page, D., Geppert, U., Zannias, T.
A\&A \textbf{360}, 1053 (2000)

\bibitem[\protect\citeauthoryear{Page et al.}{2006}]{PGW06}
Page, D., Geppert, U., Weber, F.
Nucl. Phys. \textbf{A777}, 497 (2006)

\bibitem[\protect\citeauthoryear{Page and Reddy}{2006}]{PR06}
Page, D., Reddy, S.
Annu. Rev. Nucl. Part. Sci. \textbf{56}, 327 (2006)

\bibitem[\protect\citeauthoryear{P\'e\-rez-Azor\'in et al.}{2005}]{PMP05}
P\'erez-Azor\'in, J.F., Miralles, J.A., Pons, J.A.
A\&A \textbf{433}, 275 (2005)

\bibitem[\protect\citeauthoryear{P\'e\-rez-Azor\'in et al.}{2006}]{PMP06}
P\'erez-Azor\'in, J.F., Miralles, J.A., Pons, J.A.
A\&A \textbf{451}, 1009 (2006)

\bibitem[\protect\citeauthoryear{Pons et al.}{2002}]{Petal02}
Pons, J.A., Walter, F.M., Lattimer, J.M. et al.
ApJ \textbf{564}, 981 (2002)

\bibitem[\protect\citeauthoryear{Pons}{2007}]{Pons06}
Pons, J.
in these proceedings (2007)

\bibitem[\protect\citeauthoryear{Ruderman}{2004}]{R04}
Ruderman, M.
in: Baykal, A., Yerli, S.K., Gilfanov, M., Grebenev, M. (eds)
The Electromagnetic Spectrum of Neutron Stars
[e-print: astro-ph/0410607]

\bibitem[\protect\citeauthoryear{Tr\"umper et al.}{2004}]{Tetal04}
Tr\"umper, J.E., Burwitz, V., Haberl, F. et al.
Nucl. Phys. B Proc. Suppl. \textbf{132}, 560 (2004)

\bibitem[\protect\citeauthoryear{Turolla et al.}{2004}]{TZD04}
Turolla, R., Zane, Z., Drake, J.R.
ApJ \textbf{603}, 265 (2004)

\bibitem[\protect\citeauthoryear{van Adelsberg et al.}{2005}]{vA05}
van Adelsberg M, Lai, D., Potekhin, A. Y. et al.
ApJ \textbf{628}, 902 (2005)

\bibitem[\protect\citeauthoryear{Weber}{2005}]{W05}
Weber, F.
Prog. Nucl. \& Part. Phys. \textbf{54}, 193 (2005)

\bibitem[\protect\citeauthoryear{Witten}{1984}]{W84} 
Witten, E.
Phys. Rev. \textbf{D30}, 272 (1984)

\end{thebibliography}
\end{document}